\documentclass{sig-alternate}
  \usepackage{algorithmic}
  \usepackage{algorithm}
  \usepackage{graphicx}
  \usepackage{html}

\begin{document}
\title{A Survey Paper on Recommender Systems}
\numberofauthors{6} 
\author{
% 1st. author
\alignauthor Dhoha Almazro\\
       \affaddr{Concordia University}\\
       %\affaddr{Montreal, Canada}\\
       \affaddr{d\_almaz@encs.concordia.ca}\\
      % \email{dahawee@hotmail.com}
% 2nd. author
\alignauthor Ghadeer Shahatah\\
       \affaddr{Concordia University}\\
       %\affaddr{Montreal, Canada}\\
        \affaddr{g\_shaha@encs.concordia.ca}\\
       %\email{dode\_lello@live.com}
% 3rd. author
\alignauthor Lamia Albdulkarim\\
       \affaddr{Concordia University}\\
       %\affaddr{Montreal, Canada}\\
        \affaddr{l\_alabd@encs.concordia.ca}\\
      % \email{lamiamoon@hotmail.com}
\and  % use '\and' if you need 'another row' of author names
% 4th. author
\alignauthor Mona Kherees\\
       \affaddr{Concordia University}\\
       %\affaddr{Montreal, Canada}\\
        \affaddr{m\_khere@encs.concordia.ca}\\
      % \email{mema.2@msn.com}
% 5th. author
\alignauthor Romy Martinez\\
       \affaddr{Ecole Polytechnique}\\
       %\affaddr{Montreal, Canada}\\
       \affaddr{romy.martinez@polymtl.ca}\\
% 6th. author
\alignauthor William Nzoukou\\
       \affaddr{Concordia University}\\
       %\affaddr{Montreal, Canada}\\
        \affaddr{w\_nzouko@encs.concordia.ca}\\
       %\email{nzoukou@gmail.com}
}
\maketitle
\begin{abstract}
%Finding ways of using the amount of data we collect everyday was the main idea behind data mining and knowledge discovering. For companies, one thing they can do with their historical data is to find some patterns between their products or services and their customers. Recommender systems were developed for that very purpose. They are systems that aim to predict users' interest on items (or products).
%This paper introduces the topic of recommender systems. It provides a review of some of the algorithms and techniques used to build recommender systems. Additionally, collaborative filtering
%which had become the most used technique in recommender systems and is implemented in many commercial applications will be discussed. This paper also presents an application of data mining techniques to find association between items and make predictions. We conclude by presenting one approach that could be used to improve the accuracy of the collaborative filtering algorithms.
Recommender systems apply data mining techniques and prediction algorithms to predict users' interest on information, products and services among the tremendous amount of available items. The vast growth of information on the Internet as well as number of visitors to websites add some key challenges to recommender systems. These are: producing accurate recommendation, handling many recommendations efficiently and coping with the vast growth of number of participants in the system. Therefore, new recommender system technologies are needed that can quickly produce high quality recommendations even for huge data sets. 

To address these issues we have explored several collaborative filtering techniques such as the item based approach, which identify relationship between items and indirectly \\compute recommendations for users based on these relationships. The user based approach was also studied, it identifies relationships between users of similar tastes and computes recommendations based on these relationships. 

In this paper, we introduce the topic of recommender system. It provides ways to evaluate efficiency, scalability and accuracy of recommender system. The paper also analyzes different algorithms of user based and item based techniques for recommendation generation. Moreover, a simple experiment was conducted using a data mining application -\emph{Weka}- to apply data mining algorithms to recommender system. We conclude by proposing our approach that might enhance the quality of recommender systems.
\end{abstract}
 
% A category with the (minimum) three required fields
{Clustering}

%\terms{Recommender Systems, User Based, Item Based}

\section{Introduction}
Nowadays the amount of information we are retrieving have become increasingly enormous. Back in $1982$, John Naisbitt observed that: ``we are drowning in information but starved for knowledge.'' \cite{larose}. This ``starvation'' caused by having many ways people pour data into the Internet but not many techniques to process the data to knowledge. For example, digital libraries contain tens of thousands of journals and articles. However, it is difficult for users to pick the valuable resources they want. What we really need is new technologies that can assist us find resources of interest among the overwhelming items available. 

One of the most successful such technologies is the Recommender system; as defined by M. Deshpande and G. Karypis: ``a personalized information filtering technology used to either predict whether a particular user will like a particular item (prediction problem) or to identify a set of N items that will be of interest to a certain user (top-N recommendation problem)'' \cite{4654410}.

Over the years, various approaches for building recommender systems have been created \cite{963776}; collaborative filtering has been a very successful approach in both research and practice, and in information filtering and e-commerce applications \cite{meteren}. Collaborative filtering works by creating a matrix of all items and users' preferences. In order to recommend items for the target user, similarities between him and other users are computed based on their common taste. This approach is called user-based approach. A different way to recommend items is by computing the similarities between items in the matrix. This approach is called item based approach.

%For companies (and researchers), there is an urgency to find ways to retrieve value from those information. They can do target marketing by clustering their customers %to see which category purchases a type of product. Another use of their data is to find the buying pattern of customers to recommend products. What is needed are %methods to find resources of interest. Recommender systems are ones those methods\cite{RefWorks:5}.

%Recommender systems, as defined by M. Deshpande and G. Karypis are ``personalized information filtering technology used to either predict whether a particular user will %like a particular item (prediction problem) or to identify a set of N items that will be of interest to a certain user (top-N recommendation problem)". They are used to make %predictions\footnote{How much a particular item will interest a user after he showed his interest for another item} before recommending\footnote{The system %recommend the items that have biggest prediction} items to the user. Over the years, various approaches for building recommender systems have been created %\cite{4654410}; some were based on \emph{Random prediction}, \emph{Collaborative filtering}, and \emph{Content based}.
\subsection{Background}
Recommender system can be built with many approaches. Below are some of them:
\begin{itemize} 
\item \textit{Random prediction algorithm} is an algorithm that randomly chooses items from the set of available items and recommends them to the user. Since the item's selection is done randomly, the accuracy of the algorithm is based on luck; the greater the number of items is, the chance of good selection lowers. Random prediction has a great probability of failure. Thus, it has never been taken seriously by any researcher or vendor and only serves as reference point\footnote{A similar algorithm to the Random-based algorithm is the so-called Rating-based Algorithm where instead of recommending items randomly, the system will recommend only the most popular items}, helping to compare the quality of the results obtained by the utilization of a more sophisticated algorithm \cite{Papagelis2005781}. 
\item \textit{Frequent sequences} can help build recommender systems. For example, if a customer frequently rates items we can use the frequent pattern to recommend other items to him. The only problem is that this method will only be efficient after the customer makes minimum purchases.
\item \textit{Collaborative filtering algorithms} (\emph{CF}) are algorithms that require the recommendation seekers to express their preferences by rating items. In this algorithm, the roles of recommendation seeker (a user) and preference provider\footnote{It is users who provide ratings for items} are merged; the more users rate items (or categories), the more accurate the recommendation becomes.\\ In most CF approaches, there is a list of users $U = {u_1, u_2, \ldots, u_m}$ and a list of items $I =  {i_1, i_2, \ldots, i_n}$. Each user $u_i$ has a list of item $I_{u_i}$ on which he has expressed his opinion \cite{372071}. 
\item \textit{Content based algorithms} are algorithms that attempt to recommend items that are similar to items the user liked in the past. They treat the recommendation's problem as a search for related items. Information about each item is stored and used for the recommendations. Items selected for recommendation are items that content correlates the most with the user's preferences \cite{meteren}. For example, whenever a user rated an items, the algorithm constructs a search query to find other popular items by the same author, artist, or director, or with similar keywords or subjects \cite{10.1109/MIC.2003.1167344}. Content based algorithms analyze item descriptions to identify items that are of particular interest to the user \cite{rutgers}. 
\end{itemize}
Many others approaches for recommender system exist. However, collaborative filtering algorithms have come to be the best of recommendation algorithms. As stated by Papagelis, collaborative filtering algorithms have `` been extensively adopted by both research and e-commerce recommendation systems in order to provide an intelligent mechanism to filter out the excess of information available and to provide customers with the prospect to effortlessly find out items that they will probably like according to their logged history of prior transactions'' \cite{Papagelis2005781}. \emph{CF} algorithms have significant advantages over traditional content-based filtering; they can filter any type of content, e.g. text, artwork, music, mutual funds. They can also filter based on complex and hard to represent concepts such as taste and quality. They can to make serendipitous\footnote{serendipity: the lucky tendency to find interesting or valuable things by chance (Cambridge Advanced Learner's Dictionary, 2010)} recommendations. \emph{CF} algorithms do not depend on error-prone machine analysis of content \cite{358995}.
\subsection{Contribution}
This paper has three primary research contributions:
\begin{enumerate}
\item Analysis of the user-based and item-based prediction algorithms.
\item Formulation of a hybrid model that uses both item based and user algorithm for more accurate prediction.
\item An experiment to show how from data mining we can deduce rules and make predictions.
\end{enumerate}
\subsection{Organization}
Our work will be primarily based on \emph{CF} algorithms. First, we introduce and describe collaborative filtering. Afterward, we talk two of the most used collaborative filtering algorithms; user-based and items-based algorithms. We continue by giving an example of an algorithm that is used in a commercial recommender system: the Amazon.com \emph{item-to-item} \emph{CF} algorithm. Following that, we discuss about some of the privacy and security issues related to recommender systems and also describe the metrics used to evaluate recommender. Subsequently, we continue by describing our own hybrid method. Finally we present the results of the experiments we did.

\section{Collaborative filtering algorithms}
%Collaborative filtering \emph{CF} algorithms are the most widely technique used to build recommender systems. They are algorithms that are based on historical information. 
The term ``collaborative filtering'' was first coined by Goldberg to describe an email filtering system called \emph{Tapestry\footnote{Tapestry was the first recommendation support system to be made. It was build at Xerox\textregistered Parc which also famous for inventing graphical operating system\cite{citeulike:1850360}}}. \emph{Tapestry} was an electronic messaging system that allowed users to rate messages (``good'' or ``bad'') or associate text annotations with those messages. Annotations and ratings could then be shared between users. Afterward a user could write some queries (on the annotations and on the ratings) to filter his messages. Although \emph{Tapestry} provided good recommendations, it had one major drawback; the user was required to write complicated queries \cite{963776}.  The first system to generate automated recommendations was the GroupLens\footnote{GroupLens Research is a research lab at the University of Minnesota that specialize in recommender systems.(Wikipedia , 2010) } system (Resnick et al. 1994; Konstan et al. 1997). The GroupLens system provided users with personalized recommendation on Usenet\footnote{Usenet is a worldwide distributed Internet discussion system where users read and post public messages to one or more categories, known as newsgroups. (Wikipedia 2010)} postings. It recommended articles found interesting by users similar to the target user.

Most of the \emph{CF} algorithms are based on the concept of similarity. Some algorithms (like the GroupLens system) compute the similarity between users, others look at the similarity between items, others at the similarity between categories of items. Before we can understand how \emph{CF} algorithms work, we need to understand this \emph{similarity}.
\subsection{Similarity}
Similarity (closeness) is define by data analysis in a term of a distance\footnote{the more distant objects are, the less similar they become} function such as the Euclidean (Equation \ref{eucl}) and the Manhattan (Equation \ref{manh}). 
\begin{eqnarray}
   d(i,j) &=& \sqrt{(x_{i1}-x_{j1})^2 + \ldots + (x_{in}-x_{jn})^2} \label{eucl}\\
   d(i,j) &=& |x_{i1}-x_{j1}| + \ldots + |x_{in}-x_{jn}| \label{manh}
\end{eqnarray}
In those distance functions, the difference between corresponding values of attributes in tuples $i$ and $j$ are taken. Typically attributes are normalized so that attributes with larges values do not outweigh attributes with smaller values.

The Euclidean and the Manhattan distance trivially work well and can help us compute the similarity for tuples that have attributes with numerical values. However if the attribute is categorical such as color, we need more sophisticated methods to differentiate the grading (for example color \textit{blue} vs \textit{black}) \cite{Han:dm}. Some of those methods are cosine-based similarity, Conditional Probability-Based Similarity and Pearson correlation Similarity.

\subsubsection{Cosine-Based Similarity}
\begin{figure}[h!]
\centering
\includegraphics[scale=0.70]{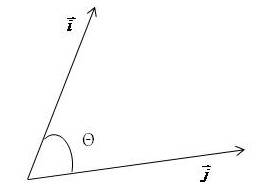}
\caption{Vector representation of items. The more $\Theta$ is small, the more similar are the items}
\label{cos}
\end{figure}

In this approach, items are thought of as vectors in the m dimensional user-space where the dimension is the attribute by which the item are rated. The cosine of the angle between the vectors that represent two items is their similarity (see Figure \ref{cos}). We know from calculus the dot--product formula: $$\vec{i}\cdot \vec{j} = ||\vec{i}||\cdot ||\vec{j}||\cdot \cos \Theta$$
$$\Longrightarrow sim(i,j) = \cos \Theta = \frac{\vec{i}\cdot \vec{j}}{||\vec{i}||\cdot ||\vec{j}||}$$
\subsubsection{Conditional Probability-Based Similarity}
Another way to compute the similarity is to use a measure that is based on the conditional probability of liking (or rating) an item given that the user already showed his interest for another item. If an item $i$ has a good chance of being purchased after an item $j$ was purchased then $i$ and $j$ are similar. The similarity is given by $sim(i,j) = P(i|j)\times \alpha$ where $\alpha$ is a factor dependent on the problem \cite{963776}.
\subsubsection{Pearson correlation Similarity}
The similarity is given by the amount of correlation between the items or users. The correlation is computed with the Pearson formula (equation \eqref{pearson}). If the set of users who both rated $i$ and $j$ are denoted by $U$ then the correlation similarity is given by:
\begin{eqnarray}
   sim(i,j) &=& corr_{ij} \label{pearson} \\ 
     & = & \frac{\sum_{u\in U}(R_{u,i}-\overline{R_i})(R_{u,j}-\overline{R_j})}{\sqrt{\sum_{u\in U}(R_{u,i}-\overline{R_i})^2}\sqrt{\sum_{u\in U}(R_{u,j}-\overline{R_j})^2}} \nonumber
\end{eqnarray}
% $$ sim(i,j) = corr_{ij} = \frac{\sum_{u\in U}(R_{u,i}-\overline{R_i})(R_{u,j}-\overline{R_j})}{\sqrt{\sum_{u\in U} R_{u,i}-\overline{R_i})^2}\sqrt{\sum_{u\in U}(R_{u,j}-\overline{R_j})^2}}$$

Experimentations have shown that Pearson correlation function performs better than cosine vector similarity (Breese et al. 1998), Spearman correlation, entropy-based uncertainty (Herlock et al. 1999). 
Pearson correlation is the most used similarity function in the two approaches of \emph{CF} based recommender; user-based or memory-based and item-based or model based \cite{963776}.
% Item based algorithms on the other hand rely on the fact that a user is mostly interested in certain items that are similar. 
\subsection{User-Based algorithms}
User based algorithms are \emph{CF} algorithms that work on the assumption that each user belongs to a group of similar behaving users. The basis for the recommendation is composed by items that are liked by users. Items are recommended based on users tastes (in term of their preference on items). The algorithm considers that users who are similar (have similar attributes) will be interested on same items \cite{lamia:3}. User based algorithms are three steps algorithm; the first step is to profile every user in order to find which ones are similar to the target user, the second step is to compute the union of the items selected by these users and associate a weight with each item based on its importance in the set and the third and final step is to select and recommend items that have the highest weight and have not been already selected by the active user \cite{963776}. The most important step is the first one; creating the union of items liked by others or selecting the most important of them is easily done when the set of similar users is known \cite{lamia:4}. Thus the overall performance of the algorithm will depend on the method used to find users that are similar to the target user. There are many methods by which it can be done. the $k$-Nearest Neighbors algorithm is the most used because of its efficiency \cite{963776}. 

$k$-Nearest Neighbors algorithm is a lazy learner classification algorithm. The algorithm requires to be provided with a training data set; a set of users who are well categorized. Then for a given user, it will compare that user's attribute with all the user in the training data set to find which ones are similar to him. The similarity between users can then be calculated using Pearson correlation (see the above sections to see why Pearson correlation is better than cosine-based and other similarity functions). %Two thoughts both based on Pearson correlation on how to find similar users exist: explicit and implicit.
Two approaches can be used to compute the similarity between users; explicitly and implicitly. 
%The explicit approach, which is based on data collection system, which is used to gather information about user such as age, user interest, demographics, etc. Moreover, user is asked to voluntarily provide their valuations, to rate some items based on assign values to an examine item. On the other hand, unlike explicit ratings in which users are asked to supply their perceptions to items explicitly in a numeric scale, implicit approach where the user is monitored based on user behavior such as transaction histories, browsing histories, product mentions, etc., are also obtainable for most e-commerce sites \cite{lamia:7}.
\subsubsection{Prediction based on explicit ratings}
In this case, users are required to express their ratings on items. This process sometimes happens through a form or a control panel. 
Let $I'={i_x:x=1,2,\ldots ,n'\wedge n'\leq n}$ where $n$ is the total number of items in the database the set of items that users $u_x$ and $u_y$ have both rated. The similarity between $u_x$ and $u_y$ is given by \cite{Papagelis2005781}:
\begin{eqnarray}
   \kappa_{x,y} & = & sim(u_x,u_y) \nonumber \\
   &=& \frac{\sum_{h=1}^{n'}(r_{u_x,i_h}-\overline{r_{u_x}})(r_{u_y,i_h}-\overline{r_{u_y}})}{\sqrt{\sum_{h=1}^{n'}(r_{u_x,i_h}-\overline{r_{u_x}})^2} \sqrt{\sum_{h=1}^{m'}(r_{u_y,i_h}-\overline{r_{u_y}})^2}}  \nonumber
\end{eqnarray}

\subsubsection{Prediction based on implicit ratings}\label{implicituser}
Implicit rating does not mean that a user will not show his appreciation toward an item, it simply means that he does not do it directly or explicitly as with the preceding approach. The rating of each item is captured implicitly. For example, if a user spend more time looking on an item, the item get an high rating. Another example is that an item will also get a high rating if an user repeatedly come look it.

\begin{flushleft}
M. Papagelisa and  D. Plexousakis define a Pearson Correlation function for a recommender where the item rating is captured by looking at the explicit rating the users gave to the categories.
\end{flushleft}

\begin{flushleft}
Let $C'={c_x:x=1,2,\ldots ,n'\wedge n'\leq n}$ where $n$ is the total number of categories in the database that users $u_x$ and $u_y$ have both rated. The similarity between $u_x$ and $u_y$ is given by \cite{Papagelis2005781}:
\end{flushleft}
\begin{eqnarray}
   \lambda_{x,y} &=& sim(u_x,u_y) \nonumber \\
   &=& \frac{\sum_{h=1}^{n'}(r_{u_x,c_h}-\overline{r_{u_x}})(r_{u_y,c_h}-\overline{r_{u_y}})}{\sqrt{\sum_{h=1}^{n'}(r_{u_x,c_h}-\overline{r_{u_x}})^2} \sqrt{\sum_{h=1}^{m'}(r_{u_y,c_h}-\overline{r_{u_y}})^2}}  \nonumber
\end{eqnarray}
where $c_x; x=1,2,\ldots ,p$ are the available categories.

After users have been clustered, the algorithms pursue by finding popular items between those users and recommend them\cite{lamia:1}.
\begin{figure}[h!]
\centering
\includegraphics[scale=0.4]{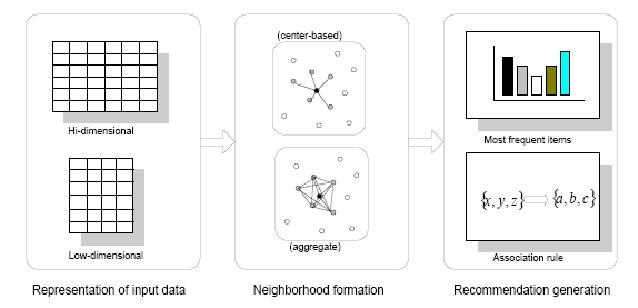}
\caption{The user-based collaborative filtering process. Image taken from \emph{Analysis of Recommendation Algorithms for ECommerce} by B. Sarwar, Ge. Karypis, J. Konstan, and J. Riedl}
\label{fig1}
\end{figure} 

%\subsubsection{User Profiling in Recommender Systems}
%To build a recommender system, recommender systems must acquire as much information as possible about user preferences. This information is generally stored in some machine understandable format called user profile. The objective of this profiling is to describe user characteristics and preferences. Thus, user profiles should contain all necessary data to model the user in recommender system \cite{lamia:8}.

%\subsubsection{User-Based Approach}
%Recommendation for a target user will be based on the opinion of other users. The user-based collaborative filtering employs statistical technique to find a set of users known as neighbors that have same preferences with the target user. Each user profile is sorted by its dissimilarity towards user profile. Selecting the top N neighbors can identify the set of similar users. Once the neighbors' list is formed, there are certain algorithms that can be applied to produce recommendation. \cite{lamia:4}
%Traditional user based approach measure the similarity and dissimilarity among users considering the whole user profile, history, interest..etc. in contrast, other developed approach measure only partial similarity between users. Two users might be very similar on one preference and completely different on other preference. Thus, on this approach, user will receive recommendation from its agent depending on other user partial similarities \cite{lamia:1}.

\subsubsection*{}
Although user based algorithms are very efficient and give good results, they suffer some drawbacks; 
\begin{itemize}
\item \textit{sparsity:}
The number of users and items in major e--commerce website is very large. Most of the users however only rated a small portion of the total items available; even very popular items result in having been rated by only a few of the total number of users. This means that the user-item matrix is very sparse and has a lot of $0$ element. Because of that it is possible that the similarity between two users cannot be defined thus making the algorithm useless \cite{citeulike:4027553}. Even when the evaluation of similarity is possible, it may not be very reliable, because of insufficient information processed.
\item \textit{scalability:}
Because finding the optimal clusters of users over large data sets is impractical. Most user-based recommenders use various forms of greedy cluster generation algorithms such as Lazy learner k-nearest neighbors. These cluster generation algorithms require a lot of computations\footnote{For example in the k-nearest neighbors algorithm, finding the optimal $k$ requires a large amount of computations} that grow linearly with the numbers of users and cannot be precomputed because users and items are changing over time in the database. Moreover, since the user based algorithms must compute the  k-nearest neighbors for every users browsing the system, the latency (waiting time) for each recommendation will increase and may it affect real-time performance of the system \cite{963776}. The conclusion is that user-based algorithm do not scale well and are not suitable for large databases of users and items.  

%\item \textit{Real-time performance:}
%Recommender systems based on user-based algorithms perform user's clustering when they start. Thus they are unable to provide recommendations for new users that are added in the database while the system is running.
\end{itemize}

\subsection{Item-Based algorithms}\label{itemb}
Because of the problems mentioned above with the user-based recommender systems, item-based (or model-based) recommender were developed. \newline
Item-based recommender are a type of \emph{collaboration} \emph{filtering} (CF) algorithms that look at the similarity between items to make a prediction. The idea is that a user is most likely to purchase items that are similar to the one he already bought in the past; so by analyzing the purchasing information we can have an idea about what he may want in the future (Deshpande, Karypis 2004). Analyzing the historical information can be done explicitly (by looking at the explicit ratings users made on the items) or implicitly (for example through the user browsing information or the rating on categories of item).  

Item-based algorithms are two steps algorithms; in the first step, the algorithms scan the past informations of the users; the ratings they gave to items are collected during this step. From these ratings, similarities between items are built and inserted into an item-to-item matrix $M$. The element $x_{i,j}$ of the matrix $M$ represents the similarity between the item in row $i$ and the item in column $j$. Afterward, in the final step, the algorithms selects items that are most similar to the particular item a user is rating. Deshpande and Karypis give a method to construct $M$ (Algorithm \ref{alg1}) after computing the similarities between the items. 
\begin{algorithm}[h!]
\caption{}
\label{alg1}
\begin{algorithmic}
  \FOR{$j \longrightarrow 1$ to $m$} 
       \FOR{$i \longrightarrow 1$ to $m$} 
             \IF{$i \neq  j$} 
                     \STATE  $M_{i,j} \longrightarrow sim(R_{*,j},R_{*,i})$
            \ELSE 
                   \STATE  $M_{i,j} \longrightarrow 0$
           \ENDIF
       \ENDFOR
       \FOR{$i \longrightarrow 1$ to $m$} 
             \IF{$i \neq  $ among the k largest values in $M_{∗, j}$} 
                     \STATE  $M_{i,j} \longrightarrow 0$
           \ENDIF
       \ENDFOR
  \ENDFOR
\end{algorithmic}
\end{algorithm}
For each item $j$, the algorithm computes the similarity between $j$ and the other items and stores the results in the $j^{th}$ column of $M$ (line 1). After that it zero-all the entries in $M$ that less similarity than the $k^{th}$ largest similarity. The second inner for-loop makes sure that an item does not recommend itself.

Similarity in item based collaborative filtering can also be computed following two approaches: implicit or explicit.
\subsubsection{Prediction based on explicit ratings}
As stated before, this approach requires users to specifically rate (give their opinion) on items.\\ Let $U' = {u_x: x= 1, 2, \ldots ,m' \wedge m' \leq m}$ where
$m$ is the total number of users in database, the set of users that have both rated item $i$ and item $j$, the Pearson correlation coefficient of their associated columns in the
user-item matrix and is given by the following formula \cite{Papagelis2005781}.
\begin{eqnarray}
   sim(i,j) &=& \frac{\sum_{h=1}^{m'}(R_{u_h,i}-\overline{R_i})(R_{u_h,j}-\overline{R_j})}   {\sqrt{\sum_{h=1}^{m'}(R_{u_h,i}-\overline{R_i})^2} \sqrt{\sum_{h=1}^{m'}(R_{u_h,j}-\overline{R_j})^2}}  \nonumber
\end{eqnarray}
$R_{u_h,i}$ is the explicit rating given by an user $u_h$ to an item $i$. And  $\overline{R_i}$ is the average of the ratings given on item $i$.
\subsubsection{Prediction based on implicit ratings}
As with the implicit user based algorithm (see section \ref{implicituser}), the ratings given to items can be implicitly captured.
 
M. Papagelisa and D. Plexousakis computes the similarity between two items as the Pearson correlation coefficient of their associated rows in the item-category bitmap matrix\footnote{``item-category bitmap matrix is a matrix of items against categories that have as elements the value 1 if the item belongs to the specific category and the value 0 otherwise.'' (M. Papagelisa, D. Plexousakis, 2005)}. 
% It is given by the following formula \cite{Papagelis2005781}:
\begin{eqnarray}
   sim(i,j) &=& \frac{\sum_{h=1}^{p}(v_{c_h,i}-\overline{v_i})(v_{c_h,j}-\overline{v_j})}   {\sqrt{\sum_{h=1}^{p}(v_{c_h,i}-\overline{v_i})^2} \sqrt{\sum_{h=1}^{p}(v_{c_h,j}-\overline{v_j})^2}}  \nonumber
\end{eqnarray}
$p$ is the number of categories and $v_{c_h,i}$ is a Boolean value that equals to $1$ if the item $i$ belongs to the category $h$ or equals to $0$ otherwise.

Compared to the user-based algorithms, item-based algorithms sparse better and scale well. Their major disadvantage is the cost to build the item-to-item matrix $M$. If we recall section \ref{itemb}, then we see that in other to construct $M$, we need to compute the similarity between every pair of items.
Once this is done, item-based algorithms perform more rapidly and scale better than the user-based algorithms. Despite their slowness, experiments have shown that user-based algorithm produce more accurate recommendation than item-based algorithms \cite{963776}. 

The choice of the algorithm will then be based on how much trade-off can be made between the prediction performance and the scalability.

\subsection{The Amazon.com example}
Amazon.com is a e-commerce website in which users can buy books, music and others goods. It has a databases containing more than 29 million customers and several million catalog items. 

Amazon.com use a algorithm based on item-based collaborative filtering to make their recommendations. Their algorithm, called \emph{item-to-item} collaborative filtering, works by first matching each of the user's purchased and rated items to similar items (as with the item based \emph{CF}, this is use to create an item-to-item matrix where elements are the similarities between items). Afterward, it combines those similar items
into a recommendation list \cite{10.1109/MIC.2003.1167344}. The most similar items are found using algorithm \ref{newalgo} (G. Linden, B.Smith, and J. York, 2003).
\begin{algorithm}[h!]
\label{newalgo}
\begin{algorithmic}
  \FOR{each item in product catalog, $I_1$} 
       \FOR{each customer $C$ who purchased $I_1$}
           \FOR{each item $I_2$ purchased by customer $C$} 
                 \STATE  Record that a customer purchased $I_1$ and $I_2$
           \ENDFOR
       \ENDFOR
       \FOR{each item $I_2$} 
            \STATE  Compute the similarity between $I_1$ and $I_2$
       \ENDFOR
  \ENDFOR
\end{algorithmic}
\caption{The most similar items algorithm. \textit{Amazon.com computes the similarity using cosine measure} }
\end{algorithm}

To improve the scalability and the performance, Amazon.com has built its recommender as two components. An offline component that creates the expensive and costly item-to-item matrix offline. The other component is the online component that look at the item-to-item matrix to produce the recommendations. The online component is  dependent only on how many titles the user has purchased or rated \cite{10.1109/MIC.2003.1167344}.
\section{Evaluation}
User satisfaction is the most important factor of the success of a recommender system which is an accurate recommendation within a reasonable time. In commercial systems, it is measured by number of recommended items that has been bought (and of course not returned!)\cite{Dhoha:7}. For non-commercial systems, it is measured by asking for users' feedback. To properly employ a recommender system, it is important to study the domain for which it is being used \cite{Dhoha:7}. 

This section will focus on evaluating recommender systems for different systems. We will then introduce three important metrics for evaluating the quality of recommender systems. Finally, we will address the challenges of employing recommender systems.
\subsection{Different Systems, Different Algorithms}
Recommender systems differ based on the type of application used. Therefore, a certain algorithm may work very well on a dataset and work poorly on different data set. In other words, some algorithms work well in situations where items are more than users (e.g. a recommender system that suggest tens of thousands of research papers to thousands of users). Other algorithms are designed for the opposite situation where users are more than items (e.g. MovieLens, a system for recommending movies, has a data set of $65000$ users and $5000$ movies) \cite{Dhoha:7}. Furthermore, Recommender Systems varies according to the nature of data sets. The static nature of items allows us to pre compute and store some of the values of the algorithm. However, the same technique is not efficient for items with a dynamic nature \cite{372071}. In some cases where similarities are way more than the dissimilarity, it is efficient to compute the dissimilarity and extract the similarity afterwards \cite{lamia:4}.
\subsection{Recommendation Metrics}
Items and users are getting increased in systems where recommender systems utilization is crucial. To ensure user satisfaction all the time, algorithms must not work on thousands, but millions of item within reasonable time \cite{Dhoha:1}.\\
Therefore, recommender systems must cope with the growth by making the suggestions more accurate, efficient and scalable.

\begin{flushleft}
\emph{Accuracy}\\
This is measured by how close the result of a recommendation matches a user's preference. Accuracy is the most important metric in evaluating the quality if a recommender system because this is what all is about: understanding the user and suggesting what he really likes or what he is looking for precisely to gain the user's trust \cite{Dhoha:1}.

There are two measures for evaluating the accuracy of a recommender system:
\begin{itemize}
\item \textbf{Statistical Accuracy Metric: } This compares the numerical recommendation scores against the actual user rating. One of the widely used metrics is the Mean Absolute Error (MAE); the lower the value of MAE the more accurate the result is \cite{372071}.
\item \textbf{Decision Support Accuracy: } which measure how effective the prediction engine is at helping a user selecting high- quality item from the set of all items. Receiver Operating Characteristic (ROC) is one of the metrics that help assessing the accuracy of predictions \cite{372071}.
\end{itemize}
An interesting point regarding accuracy was pointed out by many researchers. Very accurate recommender systems are not always good! An example is an online travel agency that recommends destinations that has already been visited by a user. Yes the recommendation was accurate enough but it was not useful she already visited these places \cite{Dhoha:4}. This means that recommendation must be accurate in predicting the upcoming actions of a user not only knowing him. Moreover, the recommendations that are most accurate according to the standards metrics are sometimes not the recommendations that are useful to users \cite{Dhoha:4}.
\end{flushleft}

\begin{flushleft}
\emph{Efficiency}\\
In order for a recommender system to be reliable, not only it must be accurate, but also it must process within a reasonable time, make good use of the available resources, and handle hundred requests per second \cite{Dhoha:1}. Memory and Computation time are two important metrics that evaluate the efficiency of a recommender system.

Algorithms that work with item sets that has a static nature tend to pre-compute item similarity and stores a matrix of similarities. The more the items, the bigger the matrix will grow. Therefore, we will end up with a quick look up table that speeds up the recommendation process; however, an $O(n^2)$ space is needed for $n$ items \cite{372071}. Because of the space problem we may not consider all the $n$ items of a system. Instead, we only consider a small fraction of the most similar items $k$ where $(k<n)$. This attempt will reduce the size of the lookup table but we will have a trade-off: smaller model size means a reduced quality \cite{372071}. Another approach to efficiently allocate space needed is to give each item a space according to the amount of rating. In other words, the more an item has ratings, the more space I allocate \cite{Dhoha:3}. 
\end{flushleft}

In some situation, the knowledge of customer preferences changes, memory consumption reduces and the time used for computation increases, therefore the efficiency of the recommender system in dynamic datasets depends on the amount of calculation required in an algorithm \cite{Dhoha:1}. In this situation, two calculations must be performed: learning time and running time. In some cases, running time was fast but learning time was 8 hours. To speed up the calculation, we consider a relevant dataset rather than the whole database; again, a trade-off between the accuracy and efficiency. Another approach to speed up the calculation is to use data structure or other data mining techniques such as hierarchical clustering since searching for neighbors is faster than scanning the whole tree \cite{Dhoha:1}.

\begin{flushleft}
\emph{Scalability}\\
A good recommendation algorithm that handles thousands of request, must also handle hundred of thousand requests in the future. Despite the accuracy and efficiency of many algorithms, they are not coping with the growth of data sets. Therefore, in order to manage the vast increase in number of users and items, a trade-off between the prediction performance and scalability is inevitable. Again, this is done by considering a portion of the whole dataset with similar characteristics.
\end{flushleft}

One of the best approaches for maintaining accuracy, efficiency, and scalability is to use hashing techniques. It compresses large data sets, scale very large number of users, and obtain a good performance within a reasonable time \cite{Dhoha:3}.
\subsection{Challenges of Recommender Systems}
If recommender systems rely only on items that have been rated, then it is missing a lot of good items for recommendation that are hidden because no one has rated them. This is called the Coverage metrics which is the percentage of items for which a recommender agent can provide predictions \cite{Papagelis2005781}. This is one of the problems that face systems that employ recommender systems. Another challenge is the sparsity issue which is rating few of the total number of items \cite{Papagelis2005781}. For systems that has just established, they are facing the cold start problem where the recommender system is unable to accurately recommend items due to the fact that only few rating has been performed on items. Noise, data redundancy, and overfitting are also other challenges of recommendation agent \cite{Dhoha:1}.

In order to reduce the sparsity problem, some researchers have proposed a compensation system by which users are rewarded for providing ratings to items. Others have proposed to capture the ratings by implicitly look at the user's behavior \cite{Sarwar98}. Another approach to solve the sparsity problem is to rely filtering agent called \textit{filterbots} or dynamic agents to automatically rate items \cite{Sarwar98}.

\section{Security and Privacy Issues}
Collaborative filtering \emph{CF} recommender requires personal information from a user to give personalized recommendations. The more users express their preferences on items, the more accurate the recommendation they receive become. As with any data mining systems, users must trust the recommender to protect their information appropriately. Moreover, since the user does not know how the recommendation is performed, he should trust the accuracy of the recommender\cite{DBLP:conf/etrics/LamFR06}. The recommender should not violate the trust of the users.

\subsection{Privacy Risks}
In most systems, users need to register before they can enjoy personalized recommendation. The registration process often requires them to provide some personal informations like their names, birth dates, postal code and email. Combinations of those required fields (attributes) may be highly identifying (\textit{Quasi-identifier\footnote{Quasi-identifier: ``Variable values or combinations of variable values within a dataset that are not structural uniques but might be empirically unique and therefore in principle uniquely identify a population unit.''(\htmladdnormallink{OECD, Glossary of statistical term}{http://stats.oecd.org/glossary/detail.asp?ID=6961}, 2010)}}). Personal preferences like those expressed to many recommender systems may become quasi-identifier, especially if some users express unusual preferences (S. Lam, D. Frankowski, and J. Riedl, 2006). User's preferences could then be used to re-identify him in another system. For example, a company like Netflix could use the preferences some users saved in its system to find them on a competitor website.

Since not every users want their information to be disclosed or misused, the recommender should then protect itself against exposition of users informations or misuse of those informations.
Recommender systems are also confronted with other type of problems such as security's related problem. Since being recommended is often promise of good selling, recommender are often target of manipulation from producers or malicious users \cite{1097061}. For example, a book writer may try to alter the recommendation so that his book get recommended. Recent research by Dellarocas and others have shown that even popular systems such as Amazon and eBay have (and are) being manipulated \cite{988726}. \emph{Shilling attacks} are one of the most discussed method by which the prediction of a recommender can be bias. 
\subsection{Shilling Attacks}
A shilling attack is an attack in which the system's recommendations for a particular item is manipulated by submitting misrepresented opinions to the system (S. Lam, D. Frankowski, and J. Riedl, 2006). The attack can have two objectives: decrease the ratings of all the items outside its target item-set (\emph{push attack}) to make them more recommended. He may also increase the ratings (\emph{nuke attack}) of other items to make its target item-set less recommended. Two simple types of shilling attacks are \emph{RandomBot} and \emph{AverageBot}.
\begin{itemize}
\item A \emph{RandomBot} is filterbot who randomly rate items outside of the target item-set with either the minimum rating (for nuke attack) or maximum rating (for push attack). 
\item An \emph{AverageBot} is a filterbot where the rating is based on the average rating of each item following a normal distribution with a mean equal to the average rating for that item. 
\end{itemize}
\begin{flushleft}
Another type of attack that may affect recommender are the so called \emph{Sybil attack} in which a dishonnest user may create multiples users account in other to improve the recommendation of another user or another item.
\end{flushleft}
\begin{flushleft}
Recommender shall then provides ways to protect itself against those attacks since they are well known. Some systems provided CAPTCHA\footnote{``A CAPTCHA or Captcha  is a type of challenge-response test used in computing to ensure that the response is not generated by a computer'' (Wikipedia, 2010)} to stop \textit{filterbots} from corrupting the ratings.
\end{flushleft}
\section{Our Approach}
After studying collaborative filtering with its both approaches item-based and user-based, we found that each approach has its advantages and disadvantages. Thus, we are proposing a new hybrid technique that combines the two approaches and trying to come up with a new approach that is more accurate and efficient.\\

The proposed approach starts by clustering all items and users based on demographic information. In other word, items and users will be categorized based on users personal attributes and make recommendations based on demographic categorization. Clustering techniques work by identifying groups of users and groups of items which appear to have similar preferences.\\

After applying the clustering technique, the next step is to extract the suitable clusters for both item based algorithm and user based algorithm. The item based algorithm will measure the similarities between a target user's preference and the items we have in the cluster. The user based algorithm will measure the similarities between the target user and other users in its cluster.\\

\begin{figure}[h!]
\centering
\includegraphics[scale=0.75]{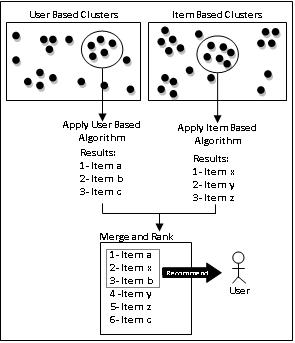}
\caption{Our hybrid approach}
\label{fig2}
\end{figure}

 The results of both algorithms are listed it terms of items. These items will be ranked from the most appropriate to the least appropriate for the target user. Then, the items in both item sets will be merged in one item set also depending on the rank that each item got in the step before. Finally, the recommendation of top-k items will be generated to the user. Figure \ref{fig2} illustrates the new hybrid approach. \\
 
 As mentioned earlier, recommender systems must cope with the growth of items and users by making suggestions more accurate, efficient and scalable. Hopefully our approach is able to handle the massive growth in a way that ensures user satisfaction.

In terms of \textbf{accuracy}, by employing item based and user based together and ranking both results, we are extracting the best of both methods and suggesting the most accurate items to users.\\

In terms of \textbf{efficiency}, the proposed approach will not going to deal with the whole database and will only deal with a portion of it due to the clustering technique that will be implemented before applying the proposed technique. Therefore, the amount of computation and memory will be much less, and it will speed up the calculation of the recommendation.\\

In terms of \textbf{scalability}, the proposed approach will not have a problem with scalability since item based algorithms is still going to be implemented and is able to handle the scalability issue. Moreover, applying a hashing technique will make the proposed system able to absorb the growth of users and items.

\section{Experiments}
We were able to implement a recommender system based on a user's profile as well as on an item based profile.  To do so, we used the Java open-source program named \emph{Weka}.   \emph{Weka} provides environment for comparing learning algorithms, graphical user interface, comprehensive set of data pre-processing tools, learning algorithms and evaluation methods.  Furthermore \emph{Weka} provides implementation of Regression, Clustering, Classification, Association rules and feature selection\footnote{\htmladdnormallink{IBM\textregistered  developerworks website}{http://www.ibm.com/developerworks/opensource/library/os-weka1/index.html}}.  As part of our experiment we used the classification algorithm \emph{J48} which is an open source Java implementation of the C4.5\footnote{C4.5 is an algorithm used to generate a decision tree that was developed by Ross Quinlan from his earlier ID3 algorithm. (\htmladdnormallink{Wikipedia}{http://en.wikipedia.org/wiki/C4.5\_algorithm},2010)} algorithm in \emph{Weka} (\htmladdnormallink{Wikipedia}{http://en.wikipedia.org/wiki/C4.5\_algorithm},2010). \emph{Weka} also provides many methods for loading data such as (ARFF) or (CSV) file, in our experiment we use a file in CSV format.

\subsection{Dataset}
In the first part of our experiment, we inputted a ``.csv'' file containing the following parameters: UserID, Age, Gender, Student, Have children, Movie category. The table below (Table \ref{schema} ) provides further details of each parameter.

In the second part , the ``.csv'' file contained the following parameters: UserID, Movie title, Movie categories: Action, Adventure, Animation, Children's, Comedy, Crime, Documentary, Drama, Fantasy, Film-Noir, Horror, Musical, Mystery, Romance, Sci-Fi, Thriller, War.  Table \ref{schema1} provides further details of each parameter.
\subsection{Data cleaning and preparation}
%Both datasets were populated using the spawner program \footnote{spawner is a program that can generates data automatically}.  After establishing which parameters we wanted to generate, we inputed them in spawner and we got some sample datasets. Each datasets contained 300 data. The movies titles in the item datasets were identical to titles provided by the MovieLens website. 
%Once we generated the datasets, user and item sets, we made sure that some entries were blank in order for weka to predict some values. Once this step was performed, the pre-processing steps consisted of four main steps: opening the file with weka, removing the attributed that we did not use in our experiment, selecting all the other attributes and finally choosing the attribute as a class attribute. These pre-processing steps were important in order to insure that the appropriate data were used in our experiments.
Both datasets were populated using the spawner\footnote{\htmladdnormallink{Spawner}{http://spawner.sourceforge.net/} is a software that can generate sample based on certain criteria} program. After establishing which parameters we wanted to generate, we inputed them in spawner and we got some sample datasets. Each datasets contained 300 data. Each datasets were modified manually in order to make sure that the dataset were coherent and logical. The movies titles in the item datasets were identical to titles provided by the MovieLens website. It is important to note that the same
movie title can be viewed and rated by different users.

Once we generated the datasets, user and item sets, we made sure that some entries were blanks in order for weka to perform some predictions. For the user data set, weka will predict a movie category whereas for the item data set, it  will predict a movie title according to which category the movie belong to. Once this step was performed, the pre-processing steps consisted of three main steps: opening the file with weka, selecting all the other attributes and finally choosing the attribute as a class attribute. These pre-processing steps were important in order to insure that the appropriate data were used in our experiments.
\subsection{Results}
%The implementation of the recommender system was done using two different input files.  Indeed, our objective was to recommend a movie title to a user according to his/her profile and also based on the movie category that movie belong to.
%Using this result (Figure \ref{tree}), we are able in the first part of the experiment, to predict what kind of movie a certain user with specific characteristics would like.  For instance, an adult, who is also a female student with children would most likely like an animation type movie.  On the other hand, a male teenager who is a student without children would rather an action movie. Therefore, we are able to recommend a certain movie category based on a user's profile. We used a user based algorithm based on demographics to generate our data.
%Using this result (Figure \ref{tree1}), we are able in the second part of the experiment, to recommend a movie title according to what type of movie category that movie belongs to.  For instance, if a user normally selects movies that are comedies, animations and a children's movies, we would recommend ``Alladin and the king of thieves''.  Furthermore, if a user rated a movie as drama and science fiction but not as a comedy and adventure, we would recommend ``Twelve Monkeys''.  Again, here we were able to recommend movie titles based on what users normally selects.
The implementation of the recommender system was done using two diffeerent input files. Indeed, our objective was to recommend a movie title to a user according to his
profile and also based on the movie category that movie belong to.

\begin{figure*}[h!]
\centering

\includegraphics[scale=0.75]{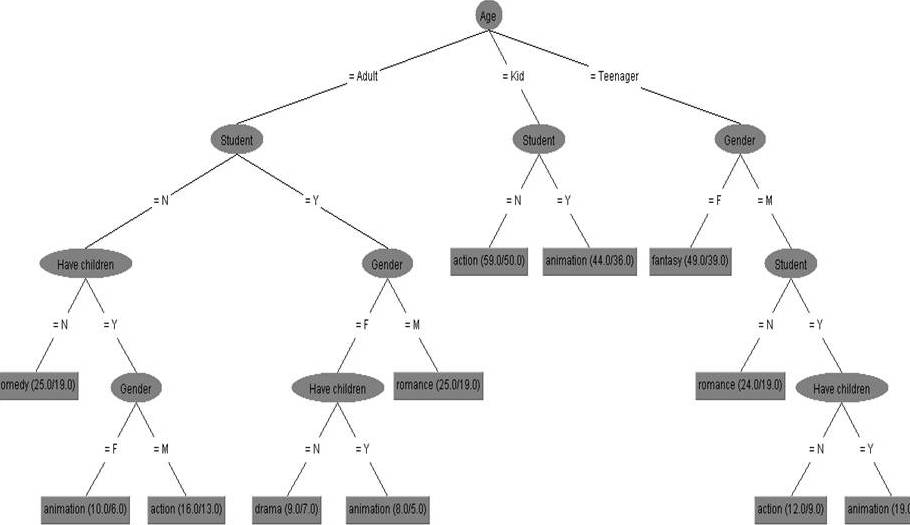}
\caption{Decision built using our training data with the schema in table \ref{schema}}
\label{tree}
\end{figure*}

%\begin{flushleft}
Using this result (Figure \ref{tree}), we are able in the first part of the experiment, to predict what kind of movie a certain user with specific characteristics would like. For instance, an adult, who is also a female student with children would most likely like an animation type movie. On the other hand, a male teenager who is a student without children would rather an action movie. Therefore, we are able to recommend a certain movie category based on a user's profile. We used a user based algorithm based on demographics to generate our data.The accuracy for this experiment was 61.43\% of correctly classified.
%\end{flushleft}

\begin{figure*}[h!]
\centering
\includegraphics[width=18.5cm, height= 23cm]{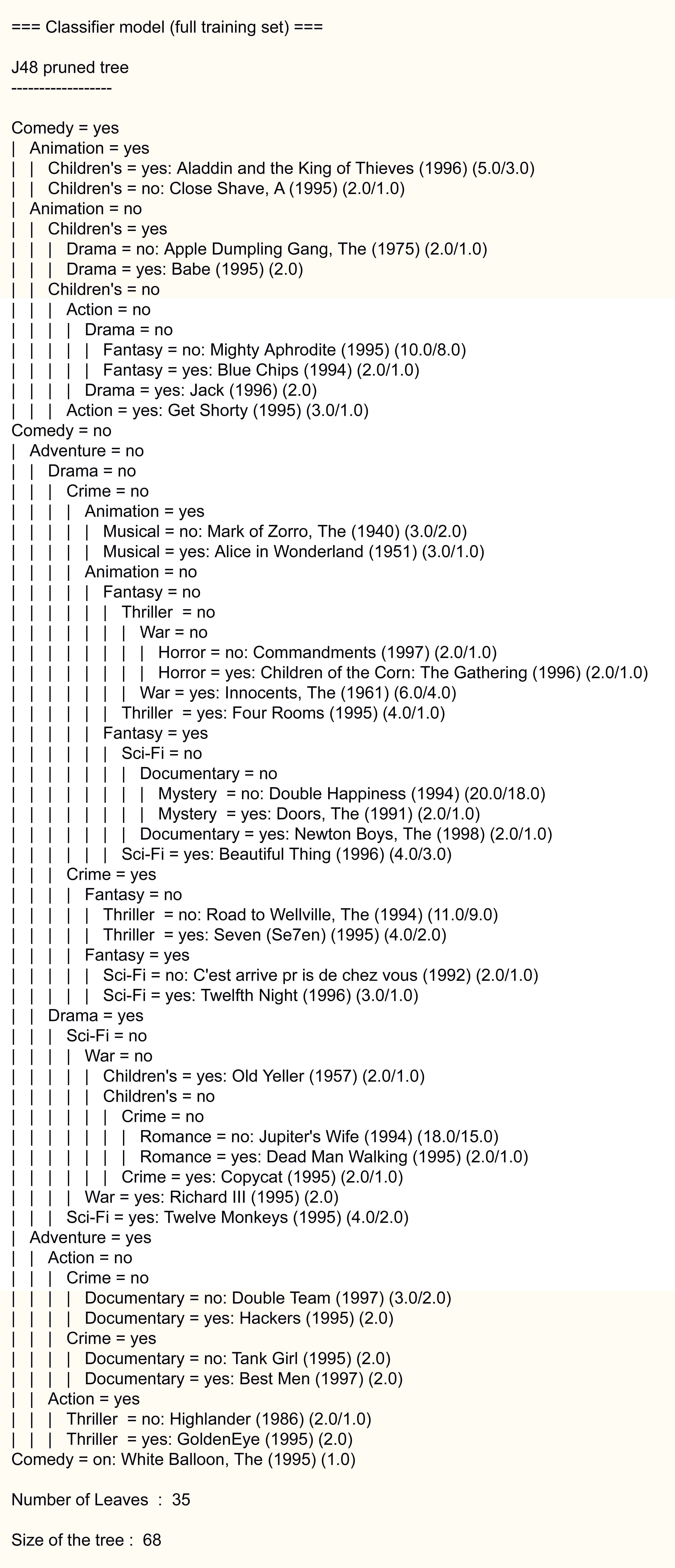}
\caption{Decision tree built using our training data with the schema in table \ref{schema1}}
\label{tree1}
\end{figure*}

%\begin{flushleft}
Using this result (Figure \ref{tree1}), we are able in the second part of the experiment, to recommend a movie title according to what type of movie category that movie belongs to. For instance, if a user normally selects movies that are comedies, animations and a children's movies, we would recommend ``Alladin and the king of thieves''. Furthermore, if a user rated a movie as drama and science fiction but not as a comedy and adventure, we would recommend ``Twelve Monkeys''. Again, here we were able to recommend movie titles based on what users normally selects. The accuracy for this experiment was 66.01\%.
%\end{flushleft}

%\begin{flushleft}
With these experiments we were able to demonstrate recommender systems using weka and trying to reproduce the user based algorithm (based on demographics) and item based algorithms.
%\end{flushleft}
\begin{table*}[h!]
\centering
\begin{tabular}{|l|p{10cm}|} \hline 
\textbf{Parameter}&\textbf{Description}\\ \hline
UserID & User who provided his favorite movie category, denoted by a numerical value\\ \hline
Age & denoted by Kid, Teenager and Adult\\ \hline
Gender & denoted by a "M" for male and "F" for female\\ \hline
Student & denoted by a``Y'' for Yes and ``N'' for No\\ \hline
Having Children & denoted by a ``Y'' for Yes and``N'' for No\\ \hline
Movie Categories & Preferred by a user Comedy, Drama, Romance, Action, Science Fiction, Crime, Documentary, Fantasy, Horror\\ \hline
\end{tabular}
\caption{Schema of the dataset we used. \textit{Note: The attribute UserID was not used in the computation of the tree.}}
\label{schema}
\end{table*}

\begin{table*}[h!]
\centering
\begin{tabular}{|l|p{10cm}|} \hline 
\textbf{Parameter}&\textbf{Description}\\ \hline
UserID & User who provided his favorite movie category, denoted by a numerical value\\ \hline
Movie title & Name of the movie\\ \hline
Movie Categories & denoted by 17 fields of categories where "yes" indicates if the movie is of that genre and "No" which indicates that it is not; movies can be in several genres at followed once.\\ \hline
\end{tabular}
\caption{Schema of the dataset}
\label{schema1}
\end{table*}

%\begin{minipage}[c]{\textwidth}
%\begin{verbatim}
%\end{verbatim}
%\end{minipage}

\section{Conclusion}
This report presented some of the algorithms used to build recommender systems. Each of these algorithm has its advantages and disadvantages; user-based algorithms are accurate but not scalable, item-based algorithms are scalable but not precise as the user-based. Research on recommender system is mainly focused on finding ways to improve the performance, scalability or accuracy of the algorithms. Hybrid algorithms that combine features of user-based and item-based algorithms have been created. Other approaches using Rough Set Prediction, Slope One Scheme Smoothing and other methods have been developed to build item-based and user-based algorithms. As with any systems that contains data on users, recommender systems have some privacy and security issues to deal with.

In conclusion, recommender systems are powerful systems that give an added-value to business and corporation. They are a relatively recent technology and they will only keep improving in the future.

%ACKNOWLEDGMENTS are optional
\section{Acknowledgments}
The authors would like to thank Dr. Benjamin Fung for 
his insightful help during the writing of this paper.\newline
This paper was produced using the ACM SAC 2010 \LaTeX$2_\epsilon$\ template.

%
% The following two commands are all you need in the
% initial runs of your .tex file to
% produce the bibliography for the citations in your paper.
\bibliographystyle{acm} %abbrv
\bibliography{references}  % references.bib is the name of the Bibliography in this case
% You must have a proper ".bib" file and remember to run:
% latex bibtex latex latex
% to resolve all references
\end{document}